\documentclass[aps,twocolumn,prb,preprintnumbers,amsmath,amssymb]{revtex4-1}
\usepackage{graphicx, bm}
\usepackage{dcolumn}
\usepackage{float}
\usepackage{latexsym}
\usepackage{amsmath}
\usepackage{graphics}
\usepackage{amssymb}
\usepackage{layout}
\usepackage{verbatim}
\usepackage{amsfonts,epsfig}
\usepackage{color}
\usepackage{setspace}
\newcommand{\beqnar}{\begin{eqnarray}}
\newcommand{\eeqnar}{\end{eqnarray}}

\newcommand{\beq}{\begin{equation}}
\newcommand{\eeq}{\end{equation}}

\begin{document}
\title{Disorder by order in graphene}
\author{S. Das Sarma, E. H.\ Hwang, and Qiuzi Li}
\affiliation{$^1$Condensed Matter Theory Center, Department of Physics, University of Maryland, College Park, Maryland 20742}
\date{\today}
\begin{abstract}
We predict the existence of an intriguing ``disorder by order" phenomenon in graphene transport where higher-quality (and thus more ordered) samples, while having higher mobility at high carrier density, will manifest more strongly insulating (and thus effectively more disordered) behavior as the carrier density is lowered compared with lower quality samples (with higher disorder) which exhibit an approximate resistivity saturation phenomenon at low carrier density near the Dirac point. This predicted behavior simulating a metal-insulator transition, which we believe to have recently been observed in an experiment at Manchester University \cite{Ponomarenko_arX11}, arises from the suppression of Coulomb disorder induced inhomogeneous puddles near the charge neutrality point in high quality graphene samples.
\end{abstract}


\maketitle

\section{Introduction}
An electronic material, metal or doped semiconductor, typically exhibits higher low-temperature conductivity as the amount of quenched random disorder is decreased in the system, i.e., as the system becomes more ordered. It is therefore a universal expectation that a purer metal with lower impurity disorder would always exhibit higher low-temperature conductivity than a dirtier metal with higher disorder.

In the current work, we theoretically establish a counter-intuitive possibility in graphene which is in sharp contrast to the universal scenario of increasing conductivity with increasing order. We show that in monolayer graphene, with its gapless chiral linear 2D electron-hole Dirac band dispersion, the resistivity (conductivity) will increase (decrease) monotonically with decreasing carrier density near the charge neutrality (Dirac) point provided the system is sufficiently pure, i.e., ordered, with very little residual background charged impurity disorder. Not only will the Dirac point resistivity be anomalously large in high-purity graphene, the transport behavior itself will be insulating-like at the charge neutrality point with the resistivity increasing monotonically with decreasing temperature! On the other hand, as the carrier density increases, the resistivity will decrease with the eventual restoration of the metallic behavior manifesting a weakly temperature dependent resistivity above a non-universal crossover density which would depend on the residual background charged impurity disorder and the temperature. At high density, far away from the charge neutrality point with vanishing average charge density, the high-purity graphene sample would behave in a perfectly normal manner manifesting very high mobility (and very long mean free path) consistent with the highly ordered nature of the system with very little residual Coulomb impurity scattering. We dub this strange dichotomy where decreasing disorder drives the graphene layer into an effective insulating state at low carrier density near the charge neutrality point, while maintaining  very high mobility at high carrier density consistent with its low disorder, the phenomenon of ``disorder by order". We emphasize that our predicted disorder by order phenomenon is not a $T=0$ quantum phase transition as in an Anderson or Mott transition, it is a transport crossover phenomenon manifesting itself as an effective density-tuned metal-insulator transition. In particular, quantum localization plays no role in our theory which is developed  entirely within the semiclassical Boltzmann transport model neglecting all quantum interference corrections. The disorder by order phenomenon arises from an interplay among charged impurity disorder, density inhomogeneity (the so-called ``electron-hole" puddles\cite{Adam_PNAS07,MCD_puddle,dassarma2010}), and the peculiar gapless linear chiral band dispersion of graphene. Our predicted novel semiclassical phenomenon would dominate low-density transport in ultrapure graphene samples as long as quantum interference induced localization corrections are small, i.e., in the effective high-temperature semiclassical regime where the inelastic phase breaking length is comparable to or smaller than the elastic transport mean free path.

We believe that our predicted graphene ``disorder by order" phenomenon has recently been experimentally observed by Ponomarenko {\it et al.} \cite{Ponomarenko_arX11}, who, however, interpret their observation as the manifestation of a density-tuned metal-insulator localization transition. All aspects of the experimental data reported in Ref.~[\onlinecite{Ponomarenko_arX11}], in particular the density and the temperature dependence of the measured conductivity, agree spectacularly well with our predictions, and we therefore contend that the observation in Ref.~[\onlinecite{Ponomarenko_arX11}] is  a direct experimental verification of our predicted ``disorder by order" phenomenon. Particularly germane in this context is the fact that the transport data of Ref.~[\onlinecite{Ponomarenko_arX11}] were taken in the relatively high-temperature ($10 - 100$ K) regime where quantum interference effects are not important, and our semiclassical transport theory should apply [See Appendix~\ref{sec:appenB} for details].

The paper is organized as follows. In Sec.~\ref{sec:tr}, the basic transport theory and the numerical results are presented.
In Section~\ref{sec:dis}, we discuss the results compared to experimental data, and we conclude in Sec.~\ref{sec:conc}.

\section{Theory and Numerical results}
\label{sec:tr}

We first provide a simple physical picture underlying the ``disorder by order" phenomenon. Let us assume that the graphene sample is pristine with essentially no background random charged impurities so that the effective transport relaxation or scattering time $\tau$  is very long, leading to very high (low)  conductivity (resistivity) at an electron density of $n$. The conductivity is given (at $T=0$) within the Boltzmann transport theory\cite{dassarma2010,HwangAdamDas_PRL07} by $\sigma = \frac{e^2 v_F^2}{2}D(E_F)\tau (E_F)$, where $v_F$ is the constant graphene Fermi velocity defining its linear band dispersion, $D(E_F) \propto \sqrt{n}/v_F$ is the graphene density of states at the Fermi energy $E_F \propto v_F \sqrt{n}$. The scattering time $\tau(E_F)$ for the screened Coulomb scattering due to random background charged impurities has been calculated \cite{dassarma2010,HwangAdamDas_PRL07,NomuraPRL,AndoMac} in the literature, giving $\tau \propto \sqrt{n}$, which leads to the now-well-known formula \cite{HwangAdamDas_PRL07} for graphene conductivity due to random charged impurity scattering given by
\begin{equation}
\sigma(n) = A \frac{n}{n_i} \frac{e^2}{h}
\label{eq:1}
\end{equation}
where $n_i$ is the effective background 2D concentration of the random charged impurities (including its location and strength) whereas $A$ is a constant which depends on the dielectric environment of the system (e.g., substrate)--for graphene on SiO$_2$ (h-BN), $A \approx 20$ (26). The result given in Eq. \ref{eq:1} and the underlying general theory for graphene carrier transport  have been well-verified experimentally in the literature\cite{ChenJ_NPH_2007}. An immediate consequence of Eq. \ref{eq:1}, interpreted naively, is that the graphene resistivity $\rho (\equiv 1/\sigma)$ diverges as $n^{-1}$ at the Dirac point where the carrier density vanishes by virtue of the vanishing density of states at the Dirac point.

In reality, however, this divergent Dirac point resistivity (or equivalently, vanishing conductivity) is not observed experimentally in real graphene samples, which manifest a conductivity saturation phenomenon at low carrier density ($|n| \lesssim n^*$\cite{Adam_PNAS07}) with an approximate minimum conductivity plateau $\sigma_{min} \sim A n^*/n_i$, where $n^*$ is a characteristic disorder-dependent density\cite{Adam_PNAS07,dassarma2010}. This conductivity minimum phenomenon was already apparent in the pioneering graphene experiments by Novoselov and Geim\cite{Novoselov}, and was later studied extensively quantitatively\cite{ChenJ_NPH_2007}, and is now accepted as arising from the charged impurity disorder induced inhomogeneous electron-hole density puddles which dominate the graphene landscape\cite{MCD_puddle} at low carrier density. These puddles of strong real-space density inhomogeneities arise from the low-density failure of screening of the individual charged impurities with electrons/holes preferably accumulating near/far from individual discrete impurities depending on the sign of the impurity charge\cite{rossi2008}. Thus, as the gate voltage decreases, the average density decreases, but electron-hole puddle formation leads to an effective saturation of the conductivity at some low sample-dependent minimum value. The inhomogeneous puddles simply cut off the $\rho \sim 1/n$ behavior of graphene resistivity for $n \lesssim n^*$ since the real 2D density across the graphene sample never vanishes although the average density does, allowing for percolating transport through the electron-hole puddles at the charge neutrality point\cite{Adam_PNAS07}.

What would happen if the electron-hole puddles are somehow eliminated or suppressed in the system? Within the semiclassical Boltzmann picture, the resistivity will become very large as the average density is decreased by lowering the gate voltage since the puddles leading to the low-density conductivity saturation phenomenon no longer exist! This is a direct (and dramatic) manifestation of the gaplessness of graphene, and cannot happen in the semiconductors with band gaps.

The easiest way to eliminate (or suppress) the puddles is, of course, to reduce the environmental charged impurity density ($n_i$) which induces the puddles to start with. But such a low-disorder system will necessarily manifest very low resistivity (since $\rho \propto n_i$) at high carrier density ($\rho \propto 1/n$), but very high resistivity near the Dirac point since $n \rightarrow 0$. If the puddles disappear completely, the resistivity will diverge as $1/n$ as the carrier density decreases. Therefore, the disorder-by-order phenomenon is peculiar to gapless graphene with its linear dispersion. It is obvious from the above physically-motivated discussion based on a qualitative extension of existing results in the literature\cite{Adam_PNAS07,dassarma2010,HwangAdamDas_PRL07,rossi2008} that this counter-intuitive ``disorder-by-order" phenomenon would be more apparent if the inhomogeneous electron-hole puddles can be further suppressed around the Dirac point  by applying an external screening potential through a gate which would screen out the puddles, as has been successfully done in Ref.~[\onlinecite{Ponomarenko_arX11}].

The above-discussed semiclassical ``disorder-by-order" phenomenon has recently been observed in the experiment of Ponomarenko {\it et al.} \cite{Ponomarenko_arX11}, who reported monotonic increase of the graphene resistivity with decreasing carrier density in an ultrapure sample on h-BN substrate. This remarkable resistivity enhancement with decreasing density occurs only in the presence of  a second nearby high-density graphene layer which screens out the puddles, thus avoiding the ``minimum-conductivity" saturation phenomenon around the Dirac point.

\begin{figure}
\includegraphics[width=0.99\columnwidth]{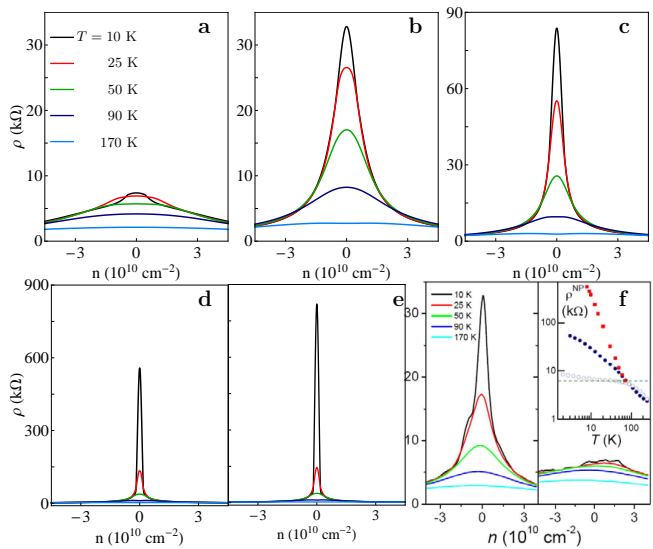}
\caption{
  Calculated $\rho(n)$ at different temperatures $T$ for $n_i = 16 \times 10^{10}$ cm$^{-2}$. (a) The potential fluctuation parameter $s= 22$ meV. (b) $s= 10$ meV. (c) $s= 6.0$ meV.    (d) $s= 1.0$ meV.  (e) $s= 0.1$ meV. (f) Experimental data as shown in Ref.~[\onlinecite{Ponomarenko_arX11}]. Comparison between (b) and (f) indicates a potential fluctuation parameter $s \sim 10$ mev in the experimental sample\cite{Ponomarenko_arX11}.
}
\label{fig:1}
\end{figure}

In Fig. \ref{fig:1} we present our theoretically calculated transport results as a function of average carrier density for the experimental situation studied by Ponomarenko {\it et al.} \cite{Ponomarenko_arX11}. The different colors in Fig. \ref{fig:1} correspond to different temperatures whereas different panels correspond to different electron-hole puddle configurations characterized by the disorder induced potential fluctuation parameter `$s$' where `$s$' corresponds to the root mean square potential fluctuations in the probability distribution function $P(V)$ for the impurity-induced disorder, $P(V) \sim e^{-V^2/2s^2}/\sqrt{2 \pi s^2}$, assumed to be Gaussian for simplicity -- the Gaussian approximation is very accurate compared with the realistic numerical calculations\cite{rossi2008,dassarma2010} of $P(V)$. The inhomogeneity parameter `$s$', which depends on the impurity disorder in the system, is directly connected to the room-mean-square density fluctuation $n_{rms}$ in the inhomogeneous electron-hole puddles. The precise relationship between $s$ and $n_{rms}$ can only be obtained through a full numerical self-consistent calculation\cite{rossi2008,dassarma2010}, but within a simple mean-field theory $n_{rms} \propto s^2$. We note that $n_{rms} \sim n^*$ defines the cut-off  for the minimum conductivity $\sigma_{min} \sim n_{rms} \sim s^2 $ around the Dirac point as discussed above. As $s \rightarrow 0$, $\rho_{CNP} = 1/\sigma_{min}$ diverges as $s^{-2}$ in the mean field theory. We note that the potential fluctuation $s$ (or equivalently, the root mean square density fluctuation in the puddles) is being controlled  by external gating through the second graphene layer in Ref.~[\onlinecite{Ponomarenko_arX11}], and the new feature of Ref.~[\onlinecite{Ponomarenko_arX11}], not achieved before, is that $s$ could be made very small.

For the sake of comparison, we have reproduced in Fig.~\ref{fig:1}(f) the corresponding experimental results from Ref.~[\onlinecite{Ponomarenko_arX11}]. The agreement between our calculated theoretical results and the experimental data is striking: In the presence of substantial (vanishing) electron-hole puddles characterized by larger (smaller) values of the disorder fluctuation parameter $s$, the calculated $\rho (n)$ saturates (increases monotonically) at lower average carrier density exactly as observed experimentally in the presence (absence) of the puddles. We emphasize that the theoretical results obtained in Fig. \ref{fig:1} use exactly the same parameters for all cases except for varying the value of the potential fluctuation parameter `$s$' as given in the figure (which mimics the effect of the suppression of $s$ through screening by the second graphene layer in Ref.~[\onlinecite{Ponomarenko_arX11}]). One interesting prediction of our theory is that $\rho (n)$ never  truly diverges as $1/n$ in our theory as $n \rightarrow 0$ (unless $s=0$ exactly, which is unphysical) since there is always a low-density cut-off $n^*$ defining the conductivity minimum regime with $n^*$ decreasing with decreasing $s$. This low-density cut-off (and the corresponding ``maximum resistivity" $\rho_{max}$) depends strongly on the puddle parameter `$s$' -- the suppression of `$s$' dramatically increases (decreases) $\rho_{max} (n^*)$.

Before discussing our results for the temperature dependence of the resistivity $\rho(T)$, we briefly discuss our transport theory \cite{Adam_PNAS07,dassarma2010,HwangAdamDas_PRL07,rossi2008,Hwang_InsuPRB2010} for graphene conductivity in the presence of electron-hole puddle induced strong density inhomogeneity. The conductivity is obtained by using the effective medium theory (EMT) by solving the integral equation
\begin{equation}
\int dn \frac{\sigma(n)-\sigma_{EMT}}{\sigma(n)+\sigma_{EMT}} P[n] = 0
\end{equation}
where $\sigma_{EMT}$ is the effective conductivity of the sample and $\sigma(n)$ is the density $n({\bm r})$ dependent local conductivity with the carrier density $n$ having the distribution $P(n)=\text{exp}[-(n-n_0)^2/2 n_{rms}^2]/\sqrt{2 \pi n_{rms}^2}$ defining the electron-hole puddles -- here $n_0$ is the average density defined by the external gate voltage (i.e., $n_0 = 0$ at the charge neutral Dirac point) and $n_{rms}$ is the root-mean-square density fluctuation due to the existence of density inhomogeneity associated with the puddles. We calculate $\sigma(n)$ using the finite-temperature Boltzmann-RPA transport theory using screened random quenched charged impurity centers (of 2D concentration $n_i$) in the environment as the resistive scattering mechanism. The Boltzmann transport theory, which has been described in details elsewhere\cite{dassarma2010,HwangAdamDas_PRL07,Hwang_InsuPRB2010,HwangScreen_PRB09} , includes five distinct temperature-dependent contributions: (1) thermal activation of electron-hole occupancy (i.e. thermal excitation of electrons from the valence band to the conduction band); (2) finite temperature thermal averaging around the Fermi surface according to the Fermi distribution function; (3) the thermal activation of carriers over the potential fluctuations associated with the electron-hole puddles; (4) finite temperature screening by the carriers themselves; (5) phonon effects (which are straightforward to include\cite{HwangDasPhonon_PRB08}, but are neglected here since electron-phonon coupling is weak in graphene). The thermal effects (1)-(3) above produce `insulating' temperature dependence, i.e., the temperature-dependent resistivity $\rho(T)$ increases with decreasing $T$,  whereas the last two effects lead to a `metallic' $\rho(T)$ decreasing with decreasing temperature. All the thermal effects are suppressed with increasing carrier density (or more precisely, increasing $E_F$), and they are the strongest at the charge neutral Dirac point (where the nominal $E_F$ vanishes).  We note that in our figures,  $\rho \equiv \sigma_{EMT}^{-1}$ whereas the density $n\equiv n_0$, i.e., the average density.


\begin{figure}
\includegraphics[width=0.76\columnwidth]{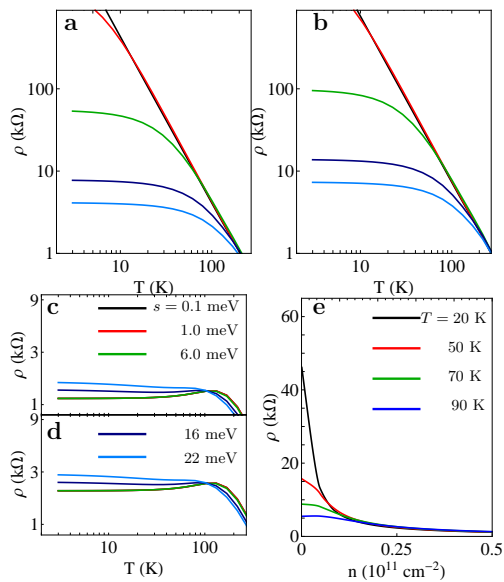}
\caption{ Calculated $\rho(T)$ for different values of the potential fluctuation parameter $s$. (a) and (b) correspond to $n = 0$, i.e., the Dirac point. (c) and (d) correspond to high carrier density $n = 5 \times 10^{10}$ cm$^{-2}$. (a) and (c) $n_i = 9 \times 10^{10}$ cm$^{-2}$ with mobility at high carrier density $\mu = 10^{5}$ cm$^2/$Vs. (b) and (d) $n_i = 16 \times 10^{10}$ cm$^{-2}$ with mobility at high carrier density $\mu = 6 \times 10^{4}$ cm$^2/$Vs. The decrease of $\rho$ for $T > 100$ K in (c) and (d) is a real effect arising from the Fermi surface averaging\cite{HwangScreen_PRB09} which becomes quantitatively important at these densities -- this effect is suppressed by phonon scattering which comes into play for $T> 100$ K. (e) $\rho(n)$ for different values of  $T$ are shown for $s= 5$ meV and $n_i = 9 \times 10^{10}$ cm$^{-2}$.
}
\label{fig:2}
\end{figure}

In Fig. \ref{fig:2} we depict our calculated $\rho(T)$ at fixed $n$ including puddle effects (characterized by the parameter $s$) within the Boltzmann transport theory as described above (and elsewhere\cite{Hwang_InsuPRB2010,HwangScreen_PRB09}). The calculated temperature dependent resistivity is identical to the experimental observations of Ref.~[\onlinecite{Ponomarenko_arX11}], as reproduced in our Fig.~\ref{fig:1}(f), with the Dirac point resistivity increasing strongly with lowering temperature for smaller values of $s$ (i.e., when the potential fluctuations associated with the electron-hole puddles are strongly suppressed) whereas the resistivity at high carrier density manifesting almost temperature-independent behavior. We emphasize, however, that within our semiclassical transport theory, in spite of the very strong increase of the Dirac point $\rho(T)$ with lowering $T$ for small $s$, eventually $\rho(T)$ saturates at some large $s$-dependent (and $n_i$-dependent) value at low enough $T \ll s$ even at the Dirac point -- empirically we find that $\rho(T)$ at the Dirac point saturates for $T \lesssim s/5$. Then, the Dirac point behavior of $\rho(T)$ for different values of $s$ is qualitatively similar if plotted as a function of $k_B T/s$. Therefore, one clear prediction of our semiclassical theory is that at low enough temperatures the experimentally measured $\rho(T)$ will always saturate even at the Dirac point, but the crossover temperature for this saturation would be very low when the electron-hole puddles are strongly suppressed. Thus, $\rho (n,T)$ in our theory would behave very much like a density-tuned metal-insulator transition, as observed in Ref.~[\onlinecite{Ponomarenko_arX11}], for all practical purposes except that the temperature dependence would be a power law (and not an exponential as in a strongly localized system), exactly as seen in Ref.~[\onlinecite{Ponomarenko_arX11}].

\section{Discussion}
\label{sec:dis}

One point we should make clear here about the ``disorder by order" phenomenon is that it arises from the suppression of the electron-hole puddles associated with the density inhomogeneity  around the Dirac point induced by Coulomb disorder, and not simply from an impurity-induced collisional broadening effect in the graphene density of states.  In particular, one may wonder whether the impurity broadening effect leading to a finite carrier density of states at the Dirac point could by itself lead to the disorder by order phenomenon since this would imply that a cleaner system would have a lower density of states at the Dirac point and hence a higher resistivity.  While this is certainly true in general, i.e. when a material exhibits a density of states minimum, addition of disorder leads to an increased density of states near this minimum due to smearing which then leads to a higher Drude conductivity, this is not the primary operational mechanism for the disorder by order phenomenon.  As we show in Appendix~\ref{sec:appenA} of this paper, such an enhanced disorder-broadened density of states indeed leads to higher conductivity near the Dirac point for more disordered graphene samples, but the conductivity always vanishes at the Dirac point ($n=0$) in this situation as long as Coulomb disorder (i.e. random charged impurities in the environment) is present in the system independent of whether the density of states is zero or finite at the Dirac point.  In our proposed order by disorder mechanism, by contrast, the conductivity is {\it always} finite at the Dirac point because of the disorder-induced inhomogeneous puddles except that this finite minimum Dirac point conductivity arising from the puddles is much lower for the more ordered systems than it is for the more disordered systems while the reverse is true at high density. Thus, the presence (suppression) of puddles is the important physics in our disorder by order mechanism, not the disorder-induced smearing of the graphene density of states around the Dirac point.  We discuss the issue of the density of states smearing effect on the graphene conductivity in Appendix~\ref{sec:appenA}.

Before concluding, we now provide a critical discussion of our theory as applied to the experimental observations of Ref. [\onlinecite{Ponomarenko_arX11}]. This issue is discussed in details in Appendix~\ref{sec:appenB}. First, our results provide an excellent description of the experimental observations with respect to the dependence of the measured resistivity as a function of carrier density, temperature, and disorder associated with the electron-hole puddles. Indeed, the agreement between our theory and the experimental data is striking, lending credence to our claim that the observation of Ref. [\onlinecite{Ponomarenko_arX11}] is an experimental verification of our predicted ``disorder by order" phenomenon in monolayer graphene\cite{FootNote}. In Ref. [\onlinecite{Ponomarenko_arX11}], the potential fluctuations associated with electron-hole puddles in the graphene layer were suppressed by a second close-by graphene layer with a very high carrier concentration. We have verified by direct numerical simulations that such a suppression can indeed be caused by the second high-density graphene layer separated by a distance $d$ acting as a gate which screens the potential fluctuations reducing them substantially below their pristine value arising from the random charged impurity distribution. A simple electrostatic analytic calculation shows that the suppression of $s$ would be approximately by a factor $1/(k_F d)$ when $k_F d \ll 1$. For $n \sim 10^{10}$ cm$^{-1}$ and $d \sim 1$ nm, $s$ could thus be suppressed by a factor as large as $50$! Thus our basic picture of the suppression of the potential fluctuation parameter `$s$' leading to the disorder by order phenomenon at the Dirac point is applicable to Ref.~[\onlinecite{Ponomarenko_arX11}]. In Ref. [\onlinecite{Ponomarenko_arX11}], it is also found that the disorder-by-order phenomenon is itself suppressed by the application of a weak magnetic field $B \sim 10$ mT. Although the inclusion of a magnetic field in the theory is beyond the scope of the current work, we mention that a magnetic field $B \sim 10$ mT corresponds to a minimum Landau level energy separation $\sim 4$ meV which is comparable to $E_F$ for $n \sim 10^9$ cm$^{-2}$. Thus, a $10$ mT magnetic field is not a weak field at the Dirac point, which would nonperturbatively modify the physics, considerably suppressing the disorder by order phenomenon.

We believe that the disorder by order phenomenon should occur in any Dirac material with chiral gapless linear energy spectrum as long as quantum interference effects are negligible, and as such, we predict the existence of the same phenomenon in the 2D surface transport\cite{CulcerFuhrer} in 3D topological insulators provided that  the surface puddles are suppressed in the system and the temperature is not too low. In fact, all gapless semiconductors will manifest the disorder by order phenomenon if impurity-induced potential fluctuations can be suppressed in the low carrier density regime. Indeed, we believe that some earlier graphene experiments \cite{BolotinDu_Sus08} observing anomalous temperature dependence of the Dirac point resistivity are observing exactly the same disorder-by-order phenomenon as reported in Ref.~[\onlinecite{Ponomarenko_arX11}] except that the authors of Ref.~[\onlinecite{BolotinDu_Sus08}] interpret their observations as ballistic transport whereas Ponomarenko {\it et al.} \cite{Ponomarenko_arX11} invoke Anderson localization!

\section{Conclusion}
\label{sec:conc}
We conclude by emphasizing that the recently observed intriguing phenomenon \cite{Ponomarenko_arX11} of monotonically increasing graphene resistivity with decreasing carrier density in graphene samples of very high purity (with very high mobilities at high carrier densities) most likely arises not from Anderson localization, but from the semiclassical ``disorder by order" phenomenon proposed in the current manuscript.  This phenomenon arises from the suppression of electron-hole puddles in the system by a near-by screening layer which then induces the system to show its intrinsic Drude behavior of the resistivity being inversely proportional to the carrier density down to much lower carrier densities without being cut off by the puddle-induced (and well-known \cite{Adam_PNAS07,dassarma2010,rossi2008,Hwang_InsuPRB2010}) ``graphene conductivity minimum" mechanism.  The qualitative difference between our ``disorder by order" mechanism and localization is that in our case, the conductivity is always finite, eventually being cut off by remnant puddles in the system at much lower carrier densities determined by the details of disorder and screening by the second layer whereas in for localization the conductivity is truly zero at $T=0$.  The other qualitative difference is that the predicted temperature dependence of the resistivity at a fixed low carrier density near the Dirac point in our mechanism is a power law whereas it must be exponential in the case of Anderson localization.  The observed temperature and density dependence of the low-density resistivity in Ref.~[\onlinecite{Ponomarenko_arX11}] is consistent with our predictions, and hence we believe that Ref.~[\onlinecite{Ponomarenko_arX11}] is manifesting the ``disorder by order" phenomenon, not Anderson localization.  When and how graphene can manifest localization (as opposed to antilocalization, which is the expected behavior for graphene) is an interesting question for the future, but we believe that the experimental temperature range must be much lower than that used ($20-100$ K) in Ref.~[\onlinecite{Ponomarenko_arX11}] to see any quantum interference induced localization effects since the inelastic phase coherence length is simply too short (Appendix~\ref{sec:appenB}) for localization effects to manifest at higher temperatures.  We urge transport experiments in high quality graphene at very low temperatures ($<1$ K) to discern localization/antilocalization versus semiclassical ``disorder by order" puddle effects.

\begin{acknowledgments}

This work is supported by US-ONR-MURI and NRI-SWAN. One of the authors (SDS) gratefully thanks Andre Geim for stimulating discussions and helpful comments.
\end{acknowledgments}

\appendix
\section{}
\label{sec:appenA}

The density of states (DOS) of disordered graphene is calculated
within the self-consistent Born
approximation (SCBA). In calculating the DOS only a short range disorder
potential is considered. The $T=0$ conductivity is calculated with a simple
formula
\begin{equation}
\sigma(n) = \frac{e^2 v_F^2}{2} D(E_F) \tau(E_F),
\label{app:sig}
\end{equation}
where $D(E_F)$ is the DOS at Fermi level, and $\tau(E_F)$ is the
transport scattering time. Note that the scattering time is calculated
with the DOS of the bare band and with two different disorders, short
range potential and long range Coulomb potential. Thus, the scattering
time is given by
\begin{equation}
\frac{1}{\tau} = \frac{1}{\tau_C} + \frac{1}{\tau_0},
\label{app:taut}
\end{equation}
where $\tau_C$ ($\tau_0$) is the scattering time due to the
long-range Coulomb impurities (short range impurities) and they are
given by in the Boltzmann transport approach
\begin{eqnarray}
\frac{1}{\tau_C} & \propto & D_b(E_F)|V_c(k_F)|^2 \propto E_F^{-1},
\nonumber \\
\frac{1}{\tau_0} & \propto & D_b(E_F)|V_0(k_F)|^2 \propto E_F,
\label{app:tau}
\end{eqnarray}
where $D_b(E) \propto E$ is the DOS of bare band.
In low density limit, $E_F \rightarrow 0$, the scattering rate by
Coulomb impurities dominates and the total scattering time
behaves as $\tau(E_F) \propto E_F$, i.e.,
$\tau(E_F) \propto n^{1/2}$. Thus, even though the DOS is finite at
$\omega=0$ the
conductivity, $\sigma(n) \propto D(E_F) \tau(E_F) \propto \sqrt{n}$ as
$n \rightarrow 0$.

\begin{figure}[t]
	\centering
	\includegraphics[width=1.0\columnwidth]{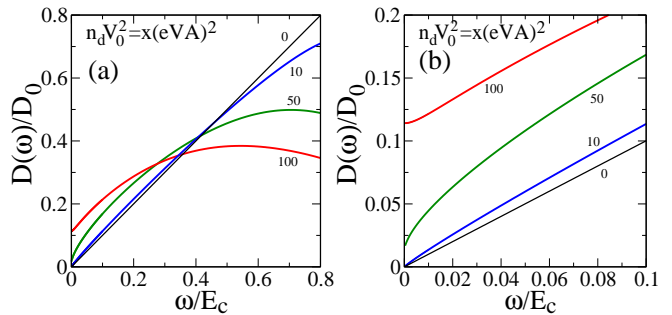}
	\caption{
Calculated the density of states within SCBA for different strengths
of  $\delta$-range disorder potential $V_0$, $n_dV_0^2 =0,$ 10, 50,
100 (eV\AA)$^2$, where $n_d$ is the short range impurity density.
Here $E_c = \hbar v_F k_c$ is the cut-off energy with $k_c \sim  1/a$ where
$a=1.42$\AA \; is the lattice constant. $D_0 = g_sg_v E_c/(2\pi (\hbar
v_F)^2)$. (b) shows the same results of (a) at low energy regime.
The DOS at $\omega=0$ is given by the formula,
$D(0)/D_0 =\ln E_0 /(2\pi E_0)$,
where $E_0=E_c/\sqrt{e^{2\pi/\gamma}-1}$ and $\gamma = n_dV_0^2/(2(\hbar v_F)^2)$.
\label{app:dos}
}
\end{figure}

\begin{figure}[t]
	\centering
	\includegraphics[width=1.0\columnwidth]{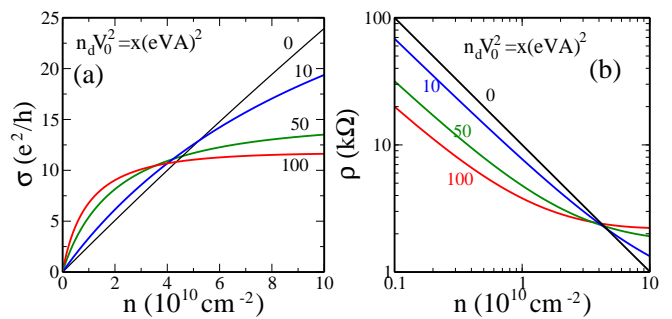}
	\caption{(color online)
Calculated conductivity as a function carrier density. Parameters of
graphene on h-BN are used. The conductivity is calculated with
Eq.~\ref{app:sig}, $\sigma = \frac{e^2v_F^2}{2} D(E_F) \tau(E_F)$,
where $D(E_F)$
is the DOS at Fermi level which is given in Fig.~\ref{app:dos}, and
$\tau(E_F)$ is the transport scattering
time calculated with DOS for the bare band. The conductivity is
calculated for a fixed long range charged
impurity $n_i = 10^{11}$ cm$^{-2}$ , but for
different short range disorder potentials as given in the figure.
Both the DOS and conductivity are calculated with the same short range
disorders. Note that the conductivity approaches zero as the carrier
density goes to zero.
(b) shows the resistivity $\rho = 1/\sigma$  calculated with the same
parameters of
(a) as a function carrier density in log-log scale.
\label{app:sig_rho}
}
\end{figure}

We calculate the disorder-broadened DOS for graphene following
Refs.~[\onlinecite{hu2008}], and
then calculate the conductivity following
Eqs.~(\ref{app:sig})--(\ref{app:tau}) above.  Our
results for the broadened DOS and the resulting conductivity are shown
in Figs.~\ref{app:dos} and \ref{app:sig_rho}.  The important
points to note are :(1) the disorder-broadened DOS leads to an
enhanced (suppressed) conductivity at low (high) carrier densities for
more (less) disordered systems in agreement with Ref.~[\onlinecite{Ponomarenko_arX11}]; (2) but the
Dirac point conductivity is always zero independent of whether the DOS
is smeared by disorder or not and thus the smearing of the DOS by
itself cannot be the explanation for the existence of the minimum
conductivity plateau in graphene which necessitates the existence of
electron-hole puddles in the system.

We can consider the DOS of disordered graphene (rather than the DOS of
the bare graphene as done above) to calculate the
scattering time. Now we have a finite DOS as $E \rightarrow 0$. Then
with the same approach as Eq.~(\ref{app:tau}) we have the scattering times
as $E_F \rightarrow 0$
\begin{eqnarray}
\frac{1}{\tau_C} & \propto & D(E_F)|V_c(k_F)|^2 \propto E_F^{-2}, \nonumber \\
\frac{1}{\tau_0} & \propto & D(E_F)|V_0(k_F)|^2 \propto E_F^{0} .
\end{eqnarray}
Similarly we have  $\tau \propto E_F^2 \propto n$, and $\sigma(n)
\propto n$ as $n \rightarrow 0$.

Thus, the DOS smearing by disorder  always produces zero conductivity
at the Dirac point ($n=0$) in graphene although close to the Dirac point
the smearing of the DOS does indeed lead to an enhanced conductivity
as shown in the figures.  Although our results are shown within the
SCBA theory of the DOS smearing, the qualitative findings are the same
within the simpler Born approximation where the disorder broadening of
the DOS does not lead to a finite DOS at zero density.  Thus, the DOS
smearing by disorder cannot be an explanation for the graphene finite
minimum conductivity around the Dirac point which arises from the
Coulomb disorder induced density inhomogeneity and electron-hole
puddles in the system.

\section{}
\label{sec:appenB}
In this Appendix, we tackle two important (and inter-related) questions: (i) What about Anderson localization? (ii) Can our Boltzmann theory be valid for $\rho > h/e^2$? We discuss these questions in some details below.

First, in the absence of inter-valley scattering graphene should manifest anti-localization\cite{TheoryAntiLocal} behavior for which there is some experimental evidence in some situations\cite{ExpAntiLocal}. The presence of inter-valley scattering would restore the usual localization behavior\cite{Aleiner_PRL}. But the crossover from the anti-localization behavior to the localization behavior should occur at an extremely low temperature comparable to the inter-valley scattering rate which is likely to be much less than $1$ K. It is therefore reasonable to assume that localization effects are operational\cite{FootNote} at much lower temperature scales than the results presented in this paper and that used in the experiments of Ref. [\onlinecite{Ponomarenko_arX11}]. In addition, the measurements of Ref. [\onlinecite{Ponomarenko_arX11}] are performed at high temperatures ($10 - 100$ K) where quantum interference effects are strongly suppressed because the inelastic phase breaking length is shorter than the elastic mean free path. The validity of the Boltzmann theory is, as a matter of principle, completely independent of the resistivity of the system as long as quantum localization/interference effects can be neglected. Thus, our semiclassical theory does not simply care about  the condition $\rho > h/e^2$ as long as quantum interference corrections are small which is true at high temperatures.

We discuss below the extent to which our semiclassical disorder-by-order transport phenomenon applies to the recent experimental observations of Ponomarenko {\it et al.} \cite{Ponomarenko_arX11}, who have interpreted their measured graphene resistivity $\rho (n, T)$ in terms of an Anderson localization driven metal-insulator transition at low carrier density.

We believe that the data of Ref.~[\onlinecite{Ponomarenko_arX11}] are inconsistent with quantum interference induced localization transition for the following reasons: (1) the temperature regime explored in \cite{Ponomarenko_arX11} is sufficiently high ($10 - 100$ K) where localization effects should be strongly suppressed since the inelastic phase breaking length is comparable to or shorter than the elastic mean free path making quantum interference irrelevant; (2) the temperature dependence of the measured resistivity in Ref.~[\onlinecite{Ponomarenko_arX11}] is not exponential and is thus inconsistent with Anderson localized insulating behavior; (3) the density dependence of the measured resistivity in Ref.~[\onlinecite{Ponomarenko_arX11}] follows our predicted semiclassical $1/n$ behavior at low density (before it saturates at a very low temperature-dependent cut-off density near the Dirac point exactly as our semiclassical theory predicts)-- this is hard to reconcile with Anderson localization; (4) there is no observable weak localization (or anti-localization) behavior manifesting in the data at high density where the resistivity is low (and weak localization corrections should be discernible if indeed a density-tuned metal-insulator strong localization phenomenon is taking place); (5) graphene can manifest true localization (even at $T=0$) only in the presence of intervalley scattering which is generally known to be extremely weak (in the absence of intervalley scattering and trigonal warping, graphene can only manifest anti-localization-- the experimental data of Ref.~[\onlinecite{Ponomarenko_arX11}] do not directly exhibit any signature of strong inter-valley scattering, for example, the high-density resistivity is completely consistent with the intervalley scattering being weak); (6) the temperature scale for the strong localization transition should be lower than the characteristic (very low) energy scale for the intervalley scattering, but the experimental data of Ref.~[\onlinecite{Ponomarenko_arX11}] manifest insulating-like behavior already at rather high ($\gg 10$ K) temperatures, casting doubts on the whole metal-insulator transition picture; and finally (7) as emphasized in our main manuscript, our semiclassical theory is in excellent agreement with the observed experimental data of Ref.~[\onlinecite{Ponomarenko_arX11}], and the observed data manifest no direct signature of quantum interference effects.

Having argued above that the experimental observations of Ponomarenko {\it et al.} \cite{Ponomarenko_arX11} are unlikely to be arising from a quantum strong localization induced metal-insulator transition, we now give the reasons for our firm belief that Ref.~[\onlinecite{Ponomarenko_arX11}] is a direct manifestation of our predicted semiclassical ``disorder by order" phenomenon: (1) our theoretical results provide an excellent quantitative description of the experimentally measured resistivity; (2) the density and the temperature dependence of the resistivity in Ref.~[\onlinecite{Ponomarenko_arX11}] follows precisely the asymptotic $\rho \sim 1/n$ and $\rho \sim 1/T^2$ behavior predicted in our theory; (3)  we predict the correct disorder dependence, i.e., lower the puddle induced inhomogeneity or potential fluctuation, stronger is the low-density insulating behavior; (4) our theory provides a clear qualitative and quantitative explanation for why the inhomogeneous electron-hole puddles must be suppressed in order to observe the apparent low-density insulating behavior; (5) our theory can explain the data both in the presence and in the absence of the puddles (with the low-density insulating behavior being respectively absent and present) whereas in the quantum localization scenario no explanation is available for the generic situation in the presence of puddles-- one must assert ad hoc that the presence of puddles somehow hinders the localization effect; (6) in the high temperature regime explored in Ref.~[\onlinecite{Ponomarenko_arX11}], quantum interference effects are strongly suppressed making our semiclassical theory applicable -- we emphasize that quantum interference cannot occur if the phase coherence length is short as it is in the temperature regime of Ref.~[\onlinecite{Ponomarenko_arX11}].

We come to the purely theoretical (rather than empirical) question of the applicability of our semiclassical theory to the experimental situation of Ref.~[\onlinecite{Ponomarenko_arX11}] where the measured resistivity is high ($\rho > h/e^2$) so that the well-known localization condition $k_F l_e >1$ (which is equivalent to the $\rho>h/e^2$ condition written out in terms of Fermi wavevector and mean free path) is violated, and the low-density insulating phase in Ref.~[\onlinecite{Ponomarenko_arX11}] actually corresponds to $k_F l_e <1$ regime, where $l_e$ is the elastic (or transport) mean free path, given by $l_e \equiv v_F \tau$, where $\tau$ is the appropriate finite-temperature elastic scattering time due to Coulomb disorder. It is indeed true that $k_F l_e \sim 1$ is approximately the condition (``the Ioffe-Regel criterion") for localization effects to become important in an electronic system with the regime $k_F l_e <1$ being dominated by quantum interference induced localization effects where semiclassical transport theory should be nominally invalid since localization corrections to the resistivity should become larger that the semiclassical resistivity itself. This is, however, true only at $T=0$ (or at extremely low temperatures) where $l_i \gg l_e$ with $l_i$ being the inelastic phase coherence length for the electrons (at $T=0$, $l_i \rightarrow \infty$, and the condition $l_i \gg l_e$ is trivially satisfied). Thus, quantum localization requires a necessary condition ($l_i \gg l_e$) so that quantum interference is operational and a sufficient condition ($\l_f > l_e$, where $l_f \sim 1/k_F$ is the Fermi wavelength). Without the necessary condition ($l_i \gg l_e$) being satisfied, quantum interference is simply not operational in a given experimental situation. (This is the reason that quantum localization phenomenon is typically experimentally studied at very low mK range of temperatures, not at $T> 10$ K.)

The necessary condition ($l_i \gg l_e$) for quantum interference (i.e., localization) to be operational is not satisfied by the experimental conditions ($T = 10 - 100$ K) used in Ref.~[\onlinecite{Ponomarenko_arX11}]. In particular, we have calculated \cite{qzli_inelastic} the inelastic phase breaking length $l_i$ due to electron-electron and electron-phonon interactions in graphene, finding that typically $l_i < l_e$ in the higher temperature ($> 20$ K) regime used in Ref.~[\onlinecite{Ponomarenko_arX11}], and even at the lowest measurement temperature in Ref.~[\onlinecite{Ponomarenko_arX11}], $T \sim 10$ K, $l_i \lesssim l_e$. Our theoretical results for $l_i$ will be published elsewhere \cite{qzli_inelastic}, but we mention that our finding that $l_i \lesssim l_e$ in the experimental regime of Ref.~[\onlinecite{Ponomarenko_arX11}] is consistent with the available direct experimental measurements of $l_i$ in graphene. For example, Ref.~[\onlinecite{Ponomarenko_arX11}] specifically quotes $l_i \sim \mu$m at liquid helium temperature ($T \sim 4$ K), which translates into $l_i \sim 10-400$ \AA \ for $T =100-20$ K since $l_i \sim T^{-2}$ in graphene due to the dominant electron-electron interaction \cite{qzli_inelastic}. The actual values of $l_i$ may even be smaller because of disorder effects and electron-phonon interactions. Direct measurements of $l_i$ in graphene \cite{Savchenko_PRL08}   are consistent with our estimate that $l_i < l_e$ in most of the temperature regime explored in Ref.~[\onlinecite{Ponomarenko_arX11}]. Thus, quantum interference phenomena are unlikely to be playing any role in Ref.~[\onlinecite{Ponomarenko_arX11}] except perhaps at the lowest temperatures ($T \sim 10$ K).

It will be desirable for the experimental regime of Ref.~[\onlinecite{Ponomarenko_arX11}] to be pushed down well below $10$ K ($T < 1$ K) where the necessary condition for quantum interference to be operational, namely, $l_i \gg l_e$, should apply. We expect our theory to fail in this low-temperature interference-dominated puddle-free regime, and it will be extremely interesting to study the deviations of the experimental data from our semi-classical theory in this low temperature ($T < 1$ K) regime so as to learn about the nature of graphene localization. We believe, for reasons discussed above, that the current experimental data of Ref.~[\onlinecite{Ponomarenko_arX11}] fall in the intriguing regime of $l_i < l_e$ and $k_F l_e \lesssim 1$ where our semiclassical theory still remains valid.

We point out a technical issue which makes our theory even more applicable to the measurements in Ref.~[\onlinecite{Ponomarenko_arX11}] than the above discussion implies. The elastic scattering length $l_e$ that $l_i$ should be compared to (so as to check if $l_i \gg l_e$ condition is satisfied for quantum interference to be relevant) is not the net mean free path defining the resistivity, but {\it only} the inter-valley scattering length $l_{iv} \gg l_e$. This is because most of the elastic resistive scattering in graphene is intravalley scattering which cannot lead to Anderson localization  (in fact, it should induce anti-localization \cite{TheoryAntiLocal}). Since the actual inter-valley scattering is very weak, according to Ref.~[\onlinecite{Ponomarenko_arX11}], $l_{iv} \sim 0.1-0.3 \mu$m, we conclude that the experimental situation in Ref.~[\onlinecite{Ponomarenko_arX11}] corresponds entirely to the regime $l_i \ll l_{iv}$ in the $T=10 - 100$ K range since $l_i \sim T^{-2}$ and therefore $l_i < 1000$ \AA \ throughout the temperature range studied in Ref.~[\onlinecite{Ponomarenko_arX11}]. Basically, the experimental conditions in Ref.~[\onlinecite{Ponomarenko_arX11}] correspond more to the infinite temperature limit than the zero temperature limit as far as quantum interference effects are concerned and as such our semiclassical theory should apply well to Ref.~[\onlinecite{Ponomarenko_arX11}]. We therefore believe that the current data of Ref.~[\onlinecite{Ponomarenko_arX11}] are well-explained by our theory to be a semiclassical disorder by order phenomenon.

We conclude by noting that in addition to the semiclassical disorder by order and the localization induced metal insulator transition phenomena, in principle there is a third possible explanation for the low-density insulating behavior observed by Ponomarenko {\it et al.} \cite{Ponomarenko_arX11}.  This is the possibility of a spontaneous interaction-induced gap formation at the Dirac point \cite{Drut_arXiv}, which would of course lead to the  insulating behavior when the chemical potential approaches the gap at low density (similar to ordinary semiconductors).  This gap scenario is unlikely to be operational in the experiment of Ref.~[\onlinecite{Ponomarenko_arX11}] because (1) typically the predicted spontaneous gap is very small, much less than the temperature range ($T=10 - 100$ K) explored in Ref.~[\onlinecite{Ponomarenko_arX11}], and (2) the temperature dependence of the resistivity in Ref.~[\onlinecite{Ponomarenko_arX11}] is a power law (roughly $T^{-2}$) which is inconsistent with the exponentially activated resistivity expected for a gapped system.  Again, experiments need to be carried out at much lower temperatures (below $1$ K) for a thorough investigation of this issue of the spontaneous gap formation, which has recently been controversial in the literature with different theories claiming both the existence and the nonexistence of an interaction-induced graphene gap formation \cite{Drut_PRL09}.


\begin{thebibliography}{10}%
\makeatletter
\providecommand \@ifxundefined [1]{%
 \ifx #1\undefined \expandafter \@firstoftwo
 \else \expandafter \@secondoftwo
\fi
}%
\providecommand \@ifnum [1]{%
 \ifnum #1\expandafter \@firstoftwo
 \else \expandafter \@secondoftwo
\fi
}%
\providecommand \enquote [1]{``#1''}%
\providecommand \bibnamefont  [1]{#1}%
\providecommand \bibfnamefont [1]{#1}%
\providecommand \citenamefont [1]{#1}%
\providecommand\href[0]{\@sanitize\@href}%
\providecommand\@href[1]{\endgroup\@@startlink{#1}\endgroup\@@href}%
\providecommand\@@href[1]{#1\@@endlink}%
\providecommand \@sanitize [0]{\begingroup\catcode`\&12\catcode`\#12\relax}%
\@ifxundefined \pdfoutput {\@firstoftwo}{%
 \@ifnum{\z@=\pdfoutput}{\@firstoftwo}{\@secondoftwo}%
}{%
 \providecommand\@@startlink[1]{\leavevmode}%
 \providecommand\@@endlink[0]{}%
}{%
 \providecommand\@@startlink[1]{%
  \leavevmode
  \pdfstartlink
   attr{/Border[0 0 1 ]/H/I/C[0 1 1]}%
   user{/Subtype/Link/A<</Type/Action/S/URI/URI(#1)>>}%
  \relax
 }%
 \providecommand\@@endlink[0]{\pdfendlink}%
}%
\providecommand \url  [0]{\begingroup\@sanitize \@url }%
\providecommand \@url [1]{\endgroup\@href {#1}{\urlprefix}}%
\providecommand \urlprefix [0]{URL }%
\providecommand \Eprint[0]{\href }%
\@ifxundefined \urlstyle {%
  \providecommand \doi [1]{doi:\discretionary{}{}{}#1}%
}{%
  \providecommand \doi [0]{doi:\discretionary{}{}{}\begingroup
  \urlstyle{rm}\Url }%
}%
\providecommand \doibase [0]{http://dx.doi.org/}%
\providecommand \Doi[1]{\href{\doibase#1}}%
\providecommand \bibAnnote [3]{%
  \BibitemShut{#1}%
  \begin{quotation}\noindent
    \textsc{Key:}\ #2\\\textsc{Annotation:}\ #3%
  \end{quotation}%
}%
\providecommand \bibAnnoteFile [2]{%
  \IfFileExists{#2}{\bibAnnote {#1} {#2} {\input{#2}}}{}%
}%
\providecommand \typeout [0]{\immediate \write \m@ne }%
\providecommand \selectlanguage [0]{\@gobble}%
\providecommand \bibinfo [0]{\@secondoftwo}%
\providecommand \bibfield [0]{\@secondoftwo}%
\providecommand \translation [1]{[#1]}%
\providecommand \BibitemOpen[0]{}%
\providecommand \bibitemStop [0]{}%
\providecommand \bibitemNoStop [0]{.\EOS\space}%
\providecommand \EOS [0]{\spacefactor3000\relax}%
\providecommand \BibitemShut [1]{\csname bibitem#1\endcsname}%
\bibitem{Ponomarenko_arX11}%
  \BibitemOpen
  \bibfield{author}{%
  \bibinfo {author} {\bibfnamefont{L.~A.}\ \bibnamefont{Ponomarenko}}, \bibinfo
  {author} {\bibfnamefont{A.~K.}\ \bibnamefont{Geim}}, \bibinfo {author}
  {\bibfnamefont{A.~A.}\ \bibnamefont{Zhukov}}, \bibinfo {author}
  {\bibfnamefont{R.}~\bibnamefont{Jalil}}, \bibinfo {author}
  {\bibfnamefont{S.~V.}\ \bibnamefont{Morozov}}, \bibinfo {author}
  {\bibfnamefont{K.~S.}\ \bibnamefont{Novoselov}}, \bibinfo {author}
  {\bibfnamefont{I.~V.}\ \bibnamefont{Grigorieva}}, \bibinfo {author}
  {\bibfnamefont{E.~H.}\ \bibnamefont{Hill}}, \bibinfo {author}
  {\bibfnamefont{V.~V.}\ \bibnamefont{Cheianov}}, \bibinfo {author}
  {\bibfnamefont{V.~I.}\ \bibnamefont{Fal'ko}}, \bibinfo {author}
  {\bibfnamefont{K.}~\bibnamefont{Watanabe}}, \bibinfo {author}
  {\bibfnamefont{T.}~\bibnamefont{Taniguchi}},\ and\ \bibinfo {author}
  {\bibfnamefont{R.~V.}\ \bibnamefont{Gorbachev}},\ }%
  \bibfield{journal}{%
  \bibinfo {journal} {Nat. Phys.}\ }%
  \textbf{\bibinfo {volume} {7}},\ \bibinfo {pages} {958} (\bibinfo {year}
  {2011})%
  \bibAnnoteFile{NoStop}{Ponomarenko_arX11}%
\bibitem{Adam_PNAS07}%
  \BibitemOpen
  \bibfield{author}{%
  \bibinfo {author} {\bibfnamefont{S.}~\bibnamefont{Adam}}, \bibinfo {author}
  {\bibfnamefont{E.~H.}\ \bibnamefont{Hwang}}, \bibinfo {author}
  {\bibfnamefont{V.~M.}\ \bibnamefont{Galitski}},\ and\ \bibinfo {author}
  {\bibfnamefont{S.}~\bibnamefont{Das~Sarma}},\ }%
  \bibfield{journal}{%
  \bibinfo {journal} {Proc.\ Natl.\ Acad.\ Sci.\ USA}\ }%
  \textbf{\bibinfo {volume} {104}},\ \bibinfo {pages} {18392 (2007); V.
  Cheianov, V. I. Fal'ko, B. L. Altshuler, and I. L. Aleiner, Phys. Rev. Lett.
  {\bf 99}, 176801} (\bibinfo {year} {2007})%
  \bibAnnoteFile{NoStop}{Adam_PNAS07}%
\bibitem{MCD_puddle}%
  \BibitemOpen
  \bibfield{author}{%
  \bibinfo {author} {\bibfnamefont{J.}~\bibnamefont{Martin}}, \bibinfo {author}
  {\bibfnamefont{N.}~\bibnamefont{Akerman}}, \bibinfo {author}
  {\bibfnamefont{G.}~\bibnamefont{Ulbricht}}, \bibinfo {author}
  {\bibnamefont{T.Lohmann}}, \bibinfo {author} {\bibfnamefont{J.~H.}\
  \bibnamefont{Smet}}, \bibinfo {author} {\bibfnamefont{K.~V.}\
  \bibnamefont{Klitzing}},\ and\ \bibinfo {author} {\bibnamefont{A.Yacoby}},\
  }%
  \bibfield{journal}{%
  \bibinfo {journal} {Nat. Phys.}\ }%
  \textbf{\bibinfo {volume} {4}},\ \bibinfo {pages} {144 (2008); Y. Zhang, V.
  W. Brar, C. Girit, A. Zettl, and M. F. Crommie, Nat. Phys. {\bf 5}, 722
  (2009); A. Deshpande, W. Bao, F. Miao, C. N. Lau, and B. J. LeRoy, Phys. Rev.
  B {\bf 79}, 205411} (\bibinfo {year} {2009})%
  \bibAnnoteFile{NoStop}{MCD_puddle}%
\bibitem{dassarma2010}%
  \BibitemOpen
  \bibfield{author}{%
  \bibinfo {author} {\bibfnamefont{S.}~\bibnamefont{Das~Sarma}}, \bibinfo
  {author} {\bibfnamefont{S.}~\bibnamefont{Adam}}, \bibinfo {author}
  {\bibfnamefont{E.~H.}\ \bibnamefont{Hwang}},\ and\ \bibinfo {author}
  {\bibfnamefont{E.}~\bibnamefont{Rossi}},\ }%
  \bibfield{journal}{%
  \bibinfo {journal} {Rev. Mod. Phys.}\ }%
  \textbf{\bibinfo {volume} {83}},\ \bibinfo {pages} {407 (2011); N. M. R.
  Peres, Rev. Mod. Phys. {\bf 82}, 2673} (\bibinfo {year} {2010})%
  \bibAnnoteFile{NoStop}{dassarma2010}%
\bibitem{HwangAdamDas_PRL07}%
  \BibitemOpen
  \bibfield{author}{%
  \bibinfo {author} {\bibfnamefont{E.~H.}\ \bibnamefont{Hwang}}, \bibinfo
  {author} {\bibfnamefont{S.}~\bibnamefont{Adam}},\ and\ \bibinfo {author}
  {\bibfnamefont{S.}~\bibnamefont{Das~Sarma}},\ }%
  \bibfield{journal}{%
  \bibinfo {journal} {Phys. Rev. Lett.}\ }%
  \textbf{\bibinfo {volume} {98}},\ \bibinfo {pages} {186806} (\bibinfo {year}
  {2007})%
  \bibAnnoteFile{NoStop}{HwangAdamDas_PRL07}%
\bibitem{NomuraPRL}%
  \BibitemOpen
  \bibfield{author}{%
  \bibinfo {author} {\bibfnamefont{K.}~\bibnamefont{Nomura}}\ and\ \bibinfo
  {author} {\bibfnamefont{A.~H.}\ \bibnamefont{MacDonald}},\ }%
  \bibfield{journal}{%
  \bibinfo {journal} {Phys. Rev. Lett.}\ }%
  \textbf{\bibinfo {volume} {98}},\ \bibinfo {pages} {076602} (\bibinfo {year}
  {2007})%
  \bibAnnoteFile{NoStop}{NomuraPRL}%
\bibitem{AndoMac}%
  \BibitemOpen
  \bibfield{author}{%
  \bibinfo {author} {\bibfnamefont{T.}~\bibnamefont{Ando}},\ }%
  \bibfield{journal}{%
  \bibinfo {journal} {J. Phys. Soc. Jpn.}\ }%
  \textbf{\bibinfo {volume} {75}},\ \bibinfo {pages} {074716} (\bibinfo {year}
  {2006})%
  \bibAnnoteFile{NoStop}{AndoMac}%
\bibitem{ChenJ_NPH_2007}%
  \BibitemOpen
  \bibfield{author}{%
  \bibinfo {author} {\bibfnamefont{J.~H.}\ \bibnamefont{Chen}}, \bibinfo
  {author} {\bibfnamefont{C.}~\bibnamefont{Jang}}, \bibinfo {author}
  {\bibfnamefont{S.}~\bibnamefont{Adam}}, \bibinfo {author}
  {\bibfnamefont{M.}~\bibnamefont{Fuhrer}}, \bibinfo {author}
  {\bibfnamefont{E.~D.}\ \bibnamefont{Williams}},\ and\ \bibinfo {author}
  {\bibfnamefont{M.}~\bibnamefont{Ishigami}},\ }%
  \bibfield{journal}{%
  \bibinfo {journal} {Nature Phys.}\ }%
  \textbf{\bibinfo {volume} {4}},\ \bibinfo {pages} {377 (2008); Y. W. Tan {\it
  et al.}, Phys. Rev. Lett. {\bf 99}, 246803 (2007); J. Heo {\it et al.}, Phys.
  Rev. B {\bf 84}, 035421 (2011); W. Zhu {\it et al.}, Phys. Rev. B {\bf 80},
  235402 (2009); C. Jang {\it et al.}, Phys. Rev. Lett. {\bf 101}, 146805
  (2008); S. Adam {\it et al.}, Phys. Rev. Lett. {\bf 101}, 046404 (2008); A.
  K. M. Newaz, {\it et al.}, Nat. Commun. {\bf 3}, 734} (\bibinfo {year}
  {2012})%
  \bibAnnoteFile{NoStop}{ChenJ_NPH_2007}%
\bibitem{Novoselov}%
  \BibitemOpen
  \bibfield{author}{%
  \bibinfo {author} {\bibfnamefont{K.~S.}\ \bibnamefont{Novoselov}}, \bibinfo
  {author} {\bibfnamefont{A.~K.}\ \bibnamefont{Geim}}, \bibinfo {author}
  {\bibfnamefont{S.~V.}\ \bibnamefont{Morozov}}, \bibinfo {author}
  {\bibfnamefont{D.}~\bibnamefont{Jiang}}, \bibinfo {author}
  {\bibfnamefont{Y.}~\bibnamefont{Zhang}}, \bibinfo {author}
  {\bibfnamefont{S.~V.}\ \bibnamefont{Dubonos}}, \bibinfo {author}
  {\bibfnamefont{I.~V.}\ \bibnamefont{Grigorieva}},\ and\ \bibinfo {author}
  {\bibfnamefont{A.~A.}\ \bibnamefont{Firsov}},\ }%
  \bibfield{journal}{%
  \bibinfo {journal} {Science}\ }%
  \textbf{\bibinfo {volume} {306}},\ \bibinfo {pages} {666 (2004); K. S.
  Novoselov, D. Jiang, F. Schedin, T. J. Booth, V. V. Khotkevich, S. V.
  Morozov, and A. K. Geim, Proc.\ Natl.\ Acad.\ Sci.\ USA {\bf 102}, 10451
  (2005); K. S. Novoselov, A. K. Geim, S. V. Morozov, D. Jiang, M. I.
  Katsnelson, I. V. Grigorieva, S. V. Dubonos, and A. A. Firsov, Nature {\bf
  438}, 197} (\bibinfo {year} {2005})%
  \bibAnnoteFile{NoStop}{Novoselov}%
\bibitem{rossi2008}%
  \BibitemOpen
  \bibfield{author}{%
  \bibinfo {author} {\bibfnamefont{E.}~\bibnamefont{Rossi}}\ and\ \bibinfo
  {author} {\bibfnamefont{S.}~\bibnamefont{Das~Sarma}},\ }%
  \bibfield{journal}{%
  \bibinfo {journal} {Phys. Rev. Lett.}\ }%
  \textbf{\bibinfo {volume} {101}},\ \bibinfo {pages} {166803 (2008); E. Rossi,
  S. Adam, and S. Das Sarma, Phys. Rev. B {\bf 79}, 245423} (\bibinfo {year}
  {2009})%
  \bibAnnoteFile{NoStop}{rossi2008}%
\bibitem{Hwang_InsuPRB2010}%
  \BibitemOpen
  \bibfield{author}{%
  \bibinfo {author} {\bibfnamefont{E.~H.}\ \bibnamefont{Hwang}}\ and\ \bibinfo
  {author} {\bibfnamefont{S.}~\bibnamefont{Das~Sarma}},\ }%
  \bibfield{journal}{%
  \bibinfo {journal} {Phys. Rev. B}\ }%
  \textbf{\bibinfo {volume} {82}},\ \bibinfo {pages} {081409 (2010); Q. Li, E.
  H. Hwang, and S. Das Sarma, Phys. Rev. B {\bf 84}, 115442} (\bibinfo {year}
  {2011})%
  \bibAnnoteFile{NoStop}{Hwang_InsuPRB2010}%
\bibitem{HwangScreen_PRB09}%
  \BibitemOpen
  \bibfield{author}{%
  \bibinfo {author} {\bibfnamefont{E.~H.}\ \bibnamefont{Hwang}}\ and\ \bibinfo
  {author} {\bibfnamefont{S.}~\bibnamefont{Das~Sarma}},\ }%
  \bibfield{journal}{%
  \bibinfo {journal} {Phys. Rev. B}\ }%
  \textbf{\bibinfo {volume} {79}},\ \bibinfo {pages} {165404 (2009); M.
  M\"uller, M. Br\"auninger, and B. Trauzettel, Phys. Rev. Lett. {\bf 103},
  196801} (\bibinfo {year} {2009})%
  \bibAnnoteFile{NoStop}{HwangScreen_PRB09}%
\bibitem{HwangDasPhonon_PRB08}%
  \BibitemOpen
  \bibfield{author}{%
  \bibinfo {author} {\bibfnamefont{E.~H.}\ \bibnamefont{Hwang}}\ and\ \bibinfo
  {author} {\bibfnamefont{S.}~\bibnamefont{Das~Sarma}},\ }%
  \bibfield{journal}{%
  \bibinfo {journal} {Phys. Rev. B}\ }%
  \textbf{\bibinfo {volume} {77}},\ \bibinfo {pages} {115449 (2008); Hongki
  Min, E. H. Hwang, and S. Das Sarma, Phys. Rev. B {\bf 83}, 161404} (\bibinfo
  {year} {2011})%
  \bibAnnoteFile{NoStop}{HwangDasPhonon_PRB08}%
\bibitem{FootNote}%
  \BibitemOpen
  \bibinfo {note} {A compelling point in this context is that the analytical
  prediction of our theory at the Dirac point, $\rho(T) \sim T^{-2}$ for $T>s$,
  is obeyed well both by our numerical results and by the experimental data of
  Ref. [\onlinecite{Ponomarenko_arX11}], arguing in favor of the disorder by
  order phenomenon being operational in the experimental observation in
  contrast to some other mechanism, e.g. localization, which typically leads to
  exponential $T$-dependence in the resistivity.}%
  \bibAnnoteFile{Stop}{FootNote}%
\bibitem{CulcerFuhrer}%
  \BibitemOpen
  \bibfield{author}{%
  \bibinfo {author} {\bibfnamefont{D.}~\bibnamefont{Culcer}}, \bibinfo {author}
  {\bibfnamefont{E.~H.}\ \bibnamefont{Hwang}}, \bibinfo {author}
  {\bibfnamefont{T.~D.}\ \bibnamefont{Stanescu}},\ and\ \bibinfo {author}
  {\bibfnamefont{S.}~\bibnamefont{Das~Sarma}},\ }%
  \bibfield{journal}{%
  \bibinfo {journal} {Phys. Rev. B}\ }%
  \textbf{\bibinfo {volume} {82}},\ \bibinfo {pages} {155457 (2010); D. Kim, S.
  Cho, N. P. Butch, P. Syers, K. Kirshenbaum, S. Adam, J. Paglione, and M. S.
  Fuhrer, Nat. Phys.} (\bibinfo {year} {in press})%
  \bibAnnoteFile{NoStop}{CulcerFuhrer}%
\bibitem{BolotinDu_Sus08}%
  \BibitemOpen
  \bibfield{author}{%
  \bibinfo {author} {\bibfnamefont{K.~I.}\ \bibnamefont{Bolotin}}, \bibinfo
  {author} {\bibfnamefont{K.~J.}\ \bibnamefont{Sikes}}, \bibinfo {author}
  {\bibfnamefont{J.}~\bibnamefont{Hone}}, \bibinfo {author}
  {\bibfnamefont{H.~L.}\ \bibnamefont{Stormer}},\ and\ \bibinfo {author}
  {\bibfnamefont{P.}~\bibnamefont{Kim}},\ }%
  \bibfield{journal}{%
  \bibinfo {journal} {Phys. Rev. Lett.}\ }%
  \textbf{\bibinfo {volume} {101}},\ \bibinfo {pages} {096802 (2008);X. Du, I.
  Skachko, A. Barker, and E. Y. Andrei, Nat. Nanotech. {\bf 3}, 491} (\bibinfo
  {year} {2008})%
  \bibAnnoteFile{NoStop}{BolotinDu_Sus08}%
\bibitem{hu2008}%
  \BibitemOpen
  \bibfield{author}{%
  \bibinfo {author} {\bibfnamefont{B.~Y.~K.}\ \bibnamefont{Hu}}, \bibinfo
  {author} {\bibfnamefont{E.~H.}\ \bibnamefont{Hwang}},\ and\ \bibinfo {author}
  {\bibfnamefont{S.~D.}\ \bibnamefont{Sarma}},\ }%
  \bibfield{journal}{%
  \bibinfo {journal} {Phys. Rev. B}\ }%
  \textbf{\bibinfo {volume} {78}},\ \bibinfo {pages} {165411 (2008); B. Dora,
  Klass Ziegler, and Peter Thalmeier, Phys. Rev. B {\bf 77}, 115422 (2008); P.
  M. Ostrovsky, I. V. Gornyi, and A. D. Mirlin, Phys. Rev. B {\bf 74}, 235443}
  (\bibinfo {year} {2006})%
  \bibAnnoteFile{NoStop}{hu2008}%
\bibitem{TheoryAntiLocal}%
  \BibitemOpen
  \bibfield{author}{%
  \bibinfo {author} {\bibfnamefont{H.}~\bibnamefont{Suzuura}}\ and\ \bibinfo
  {author} {\bibfnamefont{T.}~\bibnamefont{Ando}},\ }%
  \bibfield{journal}{%
  \bibinfo {journal} {Phys. Rev. Lett.}\ }%
  \textbf{\bibinfo {volume} {89}},\ \bibinfo {pages} {266603 (2002); E. McCann,
  K. Kechedzhi, V. I. Fal'ko, H. Suzuura, T. Ando, and B. L. Altshuler, Phys.
  Rev. Lett. {\bf 97}, 146805 (2006); A. F. Morpurgo, and F. Guinea, Phys. Rev.
  Lett. {\bf 97}, 196804 (2006); S. V. Morozov, K. S. Novoselov, M. I.
  Katsnelson, F. Schedin, L. A. Ponomarenko, D. Jiang, and A. K. Geim, Phys.
  Rev. Lett. {\bf 97}, 016801 (2006); D. V. Khveshchenko, Phys. Rev. Lett. {\bf
  97}, 036802} (\bibinfo {year} {2006})%
  \bibAnnoteFile{NoStop}{TheoryAntiLocal}%
\bibitem{ExpAntiLocal}%
  \BibitemOpen
  \bibfield{author}{%
  \bibinfo {author} {\bibfnamefont{X.}~\bibnamefont{Wu}}, \bibinfo {author}
  {\bibfnamefont{X.}~\bibnamefont{Li}}, \bibinfo {author}
  {\bibfnamefont{Z.}~\bibnamefont{Song}}, \bibinfo {author}
  {\bibfnamefont{C.}~\bibnamefont{Berger}},\ and\ \bibinfo {author}
  {\bibfnamefont{W.~A.}\ \bibnamefont{de~Heer}},\ }%
  \bibfield{journal}{%
  \bibinfo {journal} {Phys. Rev. Lett.}\ }%
  \textbf{\bibinfo {volume} {98}},\ \bibinfo {pages} {136801 (2007); F. V.
  Tikhonenko, A. A. Kozikov, A. K. Savchenko, and R. V. Gorbachev, Phys. Rev.
  Lett. {\bf 103}, 226801} (\bibinfo {year} {2009})%
  \bibAnnoteFile{NoStop}{ExpAntiLocal}%
\bibitem{Aleiner_PRL}%
  \BibitemOpen
  \bibfield{author}{%
  \bibinfo {author} {\bibfnamefont{I.~L.}\ \bibnamefont{Aleiner}}\ and\
  \bibinfo {author} {\bibfnamefont{K.~B.}\ \bibnamefont{Efetov}},\ }%
  \bibfield{journal}{%
  \bibinfo {journal} {Phys. Rev. Lett.}\ }%
  \textbf{\bibinfo {volume} {97}},\ \bibinfo {pages} {236801} (\bibinfo {year}
  {2006})%
  \bibAnnoteFile{NoStop}{Aleiner_PRL}%
\bibitem{qzli_inelastic}%
  \BibitemOpen
  \bibfield{author}{%
  \bibinfo {author} {\bibfnamefont{S.}~\bibnamefont{{Das Sarma}}}\ and\
  \bibinfo {author} {\bibfnamefont{Q.}~\bibnamefont{Li}},\ }%
  \bibinfo {note} {unpublished}%
  \bibAnnoteFile{NoStop}{qzli_inelastic}%
\bibitem{Savchenko_PRL08}%
  \BibitemOpen
  \bibfield{author}{%
  \bibinfo {author} {\bibfnamefont{F.~V.}\ \bibnamefont{Tikhonenko}}, \bibinfo
  {author} {\bibfnamefont{D.~W.}\ \bibnamefont{Horsell}}, \bibinfo {author}
  {\bibfnamefont{R.~V.}\ \bibnamefont{Gorbachev}},\ and\ \bibinfo {author}
  {\bibfnamefont{A.~K.}\ \bibnamefont{Savchenko}},\ }%
  \bibfield{journal}{%
  \bibinfo {journal} {Phys. Rev. Lett.}\ }%
  \textbf{\bibinfo {volume} {100}},\ \bibinfo {pages} {056802 (2008); D. Ki, D.
  Jeong, J. Choi, and H. Lee, Phys. Rev. B {\bf 78}, 125409} (\bibinfo {year}
  {2008})%
  \bibAnnoteFile{NoStop}{Savchenko_PRL08}%
\bibitem{Drut_arXiv}%
  \BibitemOpen
  \bibfield{author}{%
  \bibinfo {author} {\bibfnamefont{J.~E.}\ \bibnamefont{Drut}}, \bibinfo
  {author} {\bibfnamefont{T.~A.}\ \bibnamefont{L\"ahde}},\ and\ \bibinfo
  {author} {\bibfnamefont{E.}~\bibnamefont{T\"ol\"o}},\ }%
  \bibinfo {note} {arXiv:1005.5089 (2010)}%
  \bibAnnoteFile{NoStop}{Drut_arXiv}%
\bibitem{Drut_PRL09}%
  \BibitemOpen
  \bibfield{author}{%
  \bibinfo {author} {\bibfnamefont{J.~E.}\ \bibnamefont{Drut}}\ and\ \bibinfo
  {author} {\bibfnamefont{T.~A.}\ \bibnamefont{L\"ahde}},\ }%
  \bibfield{journal}{%
  \bibinfo {journal} {Phys. Rev. Lett.}\ }%
  \textbf{\bibinfo {volume} {102}},\ \bibinfo {pages} {026802 (2009); J. Wang,
  and G. Liu, New J. Phys. 14, 043036} (\bibinfo {year} {2012})%
  \bibAnnoteFile{NoStop}{Drut_PRL09}%
\end{thebibliography}
\end{document}